\documentstyle[preprint,aps,epsf]{revtex}
\tighten

\begin{document}
\newcommand{\pdwa}{^{\prime\prime}}
\newcommand{\ptrzy}{^{\prime\prime\prime}}       

\draft

\title{Different formulations of $^3$He and $^3$H photodisintegration
}

\author{
R.~Skibi\'nski$^1$,
J.~Golak$^1$,
H.~Wita\l{}a$^1$,
W.~Gl\"ockle$^2$,
A.~Nogga$^3$.
}
\address{$^1$M. Smoluchowski Institute of Physics, Jagiellonian University,
                    PL-30059 Krak\'ow, Poland}
\address{$^2$Institut f\"ur Theoretische Physik II,
         Ruhr-Universit\"at Bochum, D-44780 Bochum, Germany}
\address{$^3$ Institute for Nuclear Theory, University of Washington,
                  Box 351550 Seattle, WA 98195, USA}

\date{\today}
\maketitle

\begin{abstract}
Different momentum space Faddeev-like equations and their solutions
for the radiative pd-capture and the three-nucleon
photodisintegration of $^3$He are presented.
Applications are based on
the AV18 nucleon-nucleon and the Urbana IX three nucleon forces.
Meson exchange currents are included using the Siegert theorem.
A very good agreement has been found in all cases indicating the reliability of the used
numerical methods.
Predictions for cross sections and polarization observables in the pd-capture
and the complete three nucleon breakup of $^3$He at different incoming deuteron/photon energies
 are presented.  
\end{abstract}

\narrowtext

\section{Introduction}
\label{secIN}

In the case of few nucleon systems it is nowadays possible to compare precise experimental data
and theoretically well controlled predictions.
For low energy processes with three nucleons (3N), theoretical results can be obtained for any given
realistic nuclear interaction. This makes such systems
an important tool for the investigation of the nuclear Hamiltonian.
In momentum space the formalism of Faddeev
equations has been used to obtain bound and scattering nuclear
states~\cite{ref.Report}.
A very good agreement between theoretical predictions and experimental data was obtained (e.g.~\cite{kistryn}).
An equivalent description was also obtained
using configuration space variational methods~\cite{kievsky}.
After such encouraging  results in pure nucleonic systems had become available also weak and electromagnetic
processes with three nucleons were studied in the same scheme.
In a series of papers we
showed the results for electron induced processes~\cite{golakelek}, proton-deuteron radiative
capture~\cite{Golak.pdc.2000}, muon capture~\cite{muon} and 3N bound states photodisintegration~\cite{skib2b,skib3b}.
It was found that all dynamical ingredients are important:
final state interactions play a significant role and the clear effect of 3N forces
is noticed.
On top of that also the addition of meson exchange currents changes predictions
in a significant way, making the analysis based only on the single nucleon current mostly
meaningless.
Inclusion of all those components allows for precise predictions on different observables,
like cross sections or asymmetries and consequently the door is open for
investigations of other physical issues
e.g. neutron electromagnetic formfactors~\cite{form}.
 
For very low energies hyperspherical harmonic expansion methods were used by the Pisa group~\cite{pisa}
and a nice agreement with our results was observed.
The total photodisintegration cross sections were also calculated by the Trento group~\cite{trento} using
the Lorentz integral transform method. We compared our results in a common paper~\cite{trento.benchmark}.
Also the Hanover group presented results on photo-~\cite{hannover} and
electro-~\cite{hannover.ele} disintegrations.
In their approach the $\Delta$ degree of freedom and corresponding nuclear
currents are taken explicitly into account. These  predictions are also in qualitative agreement with ours.  

In the Faddeev scheme different formulations of the three-body problem are possible.
In this work we would like to compare different ways of obtaining transition amplitudes
for two- and three- body photodisintegration of 3N bound states.
The comparison between predictions based on the different formulations is
a strong test of the used numerical methods.
The numerical calculation of the three body continuum is a non-trivial problem
and the possibility of testing different formulations deserves thorough investigations.
Up to now even without 3NF's, no direct systematic comparison of exclusive three-body
photodisintegration cross sections between different approaches was performed.
Including 3NF's the situation is even worse.
To the best of our knowledge, no other collaboration has done calculations with explicit
3NF's. Therefore, an internal comparison is of utmost importance.
It also provides useful information on the efficiency of the different
formulations for practical calculations.
 
The calculations presented in this paper are based on the AV18 NN potential~\cite{ref.AV18} alone,
and combined with the Urbana IX 3N force~\cite{urbana}.
Since the scope of this paper is not an investigation of details of the electromagnetic current operator,
the same model of the current is used in all investigated formulations.
The single nucleon current is
supplemented by some exchange currents included by the Siegert theorem.
This approach is described in~\cite{Golak.pdc.2000} in more detail.
 
 
 
In Section II we describe three ways of obtaining the transition amplitude for the radiative Nd-capture
(and equivalently for the two-body photodisintegration of the 3N bound state) and
two methods for the three-body photodisintegration of the 3N bound state.
In Section III we compare predictions based on those methods.
We summarize in Section IV.                                        

\section{Theoretical Framework}
\label{secII}

In this section we would like to describe three different methods used to
generate transition amplitudes for the Nd-capture and the two-body photodisintegration
of the 3N bound state.
We also show how to build the transition amplitudes for the three-body photodisintegration.
 
The nuclear matrix element for the two-body photodisintegration
of the 3N bound state $\mid \Psi_b \rangle$ is
\begin{equation}
N^{Nd}_{\mu} \equiv \langle \Psi_{Nd}^{(-)} \mid j_{\mu} \mid \Psi_b \rangle \;,
\end{equation}
where $\langle \Psi_{Nd}^{(-)} \mid$ is the final scattering state.
In the Faddeev scheme it can be presented in the form~\cite{skib2b}
\begin{equation}
N^{Nd}_{\mu} = \langle \psi_1 \mid (1+P) j_{\mu} \mid \Psi_b \rangle\;,
\end{equation}
where $\langle \psi_1 \mid$
is a Faddeev component of $\mid \Psi_{Nd}^{(-)} \rangle$ and
$j_{\mu}$ is the electromagnetic current operator.
$P$ is a permutation operator defined as a sum of cyclical and anti-cyclical
permutations of three particles
\begin{equation}
P \equiv P_{12}P_{23} + P_{13}P_{32},
\end{equation}
where $P_{ij}$ interchanges
the i-th and j-th nucleons.
 
The Faddeev amplitude $\langle \psi_1 \mid$ obeys the Faddeev-like equation~\cite{huber}
\begin{eqnarray}
\langle \psi_1 \mid &=& \langle \phi_1 \mid + \langle \psi_1 \mid [ Pt_1G_0
+ (1+P)V_4^{(1)}G_0(t_1G_0+1) ] \nonumber \\
&\equiv& \langle \phi_1 \mid + \langle \psi_1 \mid K\;,
\end{eqnarray}
where $\mid \phi_1 \rangle$ is a product of the deuteron state
and a momentum eigenstate of the spectator nucleon.
Further, $V_4^{(1)}, G_0$ and $t_1$ are a part of the 3NF symmetrical under exchanges of nucleons 2 and 3,   
the free three-nucleon propagator and the
two-body t-operator acting in the 2-3 subspace, respectively.
Thus
\begin{equation}
\langle \psi_1 \mid = \langle \phi_1 \mid (1-K)^{-1}
\end{equation}
and
\begin{equation}
N^{Nd}_{\mu} = \langle \phi_1 \mid (1-K)^{-1} (1+P) j_{\mu} \mid \Psi_b \rangle\,.
\end{equation}
Defining the auxiliary state $\mid U \rangle$
\begin{equation}
\mid U \rangle \equiv (1-K)^{-1} (1+P) j_{\mu} \mid \Psi_b \rangle \label{defU}
\end{equation}
one gets
\begin{equation}
N^{Nd}_{\mu} = \langle \phi_1 \mid U \rangle \;. \label{neweq8}
\end{equation}
According to the definition~(\ref{defU}) the state $\mid U \rangle$ fulfills
\begin{equation}
\mid U \rangle = (1+P)j_{\mu} \mid \Psi_b \rangle + K \mid U \rangle\;.
\end{equation}
Inserting $K$ this reads
\begin{eqnarray}
\mid U \rangle &=& (1+P)j_{\mu} \mid \Psi_b \rangle \nonumber \\
&+& [ Pt_1G_0 + (1+P)V_4^{(1)}G_0(t_1G_0+1) ] \mid U \rangle\;.
\label{Eq:start}
\end{eqnarray}
This form of the kernel with $P$ standing to the left causes unnecessary complications
since the deuteron pole in $t_1$ appears as smeared-out into a logarithmic singularity\cite{ref.Report}.
We avoid that by reformulation of
Eq.(~\ref{Eq:start}), as is shown below.
 
\subsection{Methods $1_{NN+3NF}$ and $1_{NN}$}         
Denoting $\mid \chi \rangle \equiv (1+P)j_{\mu} \mid \Psi_b \rangle$ and
introducing the auxiliary states $\mid U' \rangle$ and
$\mid U\pdwa \rangle$:
\begin{eqnarray}
\mid U' \rangle &\equiv& t_1G_0 \mid U \rangle \\
\mid U\pdwa \rangle  &\equiv& V_4^{(1)}G_0(t_1G_0+1) \mid U \rangle \label{met10}
\end{eqnarray}
one gets
\begin{eqnarray}
N^{Nd}_{\mu} &=& \langle \phi_1 \mid \chi \rangle + \langle \phi_1 \mid
 P \mid U' \rangle \nonumber \\
&+& \langle \phi_1 \mid (1+P) \mid U\pdwa \rangle \;.\label{r6}
\end{eqnarray}
 
The states $\mid U' \rangle$ and
$\mid U\pdwa \rangle$ fulfill the set of coupled equations:
 
\begin{eqnarray}
\mid U' \rangle &=& t_1G_0 (1+P) j_{\mu} \mid \Psi_b \rangle +
t_1G_0P \mid U' \rangle \nonumber \\
&+& t_1G_0 (1+P) \mid U\pdwa \rangle \label{met11}  \\
\mid U\pdwa \rangle &=& V_4^{(1)}G_0(1+t_1G_0)\mid (1+P) j_{\mu} \mid \Psi_b \rangle \nonumber \\
&+& V_4^{(1)}G_0(1+t_1G_0)P \mid U' \rangle \nonumber \\
&+& V_4^{(1)}G_0(1+t_1G_0) (1+P) \mid U\pdwa \rangle \;. \label{met11a}
\end{eqnarray}
 
Solving numerically the set of Eqs~(\ref{met11})-(\ref{met11a}) and using Eq.~(\ref{r6}) one gets
the transition amplitude $N^{Nd}_{\mu}$.
In the following, the results based on the Eqs~(\ref{r6})-(\ref{met11a})
will be denoted as "method $1_{NN+3NF}$."
 
In the case when only the NN interaction is used ($V_4^{(1)}=0 \Rightarrow \mid U\pdwa \rangle=0$)
Eq.~(\ref{met11}) simplifies to
\begin{equation}
\mid U' \rangle = t_1G_0 (1+P) j_{\mu} \mid \Psi_b \rangle +
t_1G_0P \mid U' \rangle \label{met}
\end{equation}
and
\begin{equation}
N^{Nd}_{\mu} = \langle \phi_1 \mid \chi \rangle + \langle \phi_1 \mid
 P \mid U' \rangle
\end{equation}
 
This will be called "method $1_{NN}$."    
 
In all our methods the inhomogeneous integral equations will always be solved
by iteration and consecutive Pad\'e summation.
In the numerical implementation it is important that in both methods, during the iterations
of the set of Eqs.~(\ref{met11})-(\ref{met11a}) or Eq.~(\ref{met}) the permutation
operators from the integral kernels act only onto the
$t-$operator or the 3NF's matrix elements,
which stand at the very left in the driving terms of Eqs.(~\ref{met11})-(\ref{met}).
We work in a partial wave decomposition and the presence of the
nuclear interactions (in $t_1$ and $V_4$)
enforces that only channels with relatively small partial waves are important.
Thus the $P$ operator which acts upon
$t_1$ or $V_4$ can also be taken using a relatively small number of partial waves.
 
Having solved the set of Eqs.~(\ref{met11})-(\ref{met11a}) one can also obtain the
amplitude for three-body
photodisintegration~\cite{skib2b}
\begin{equation}
N^{3N}_{\mu} = \langle \phi_{3N} \mid (1+P) j_{\mu} \mid \Psi_b \rangle  +
\langle \phi_{3N} \mid (1+P) \lbrace \mid U' \rangle + \mid U\pdwa \rangle \rbrace \;.
\end{equation}
Here $\langle \phi_{3N} \mid$ is a product of momentum eigenstate describing
three free nucleons and
antisymmetrized in the 23 subsystem.
 
\subsection{Methods $2_{NN+3NF}$ and $2_{NN}$}  
The second method, which we will denote as "$2_{NN+3NF}$" has been presented in detail in~\cite{skib2b}, where
also some predictions for the pd-capture and the two body photodisintegration were shown.
However, for the purpose of completeness we briefly describe also this method.
Using the identity
\begin{equation}
  1 + P  = \frac12 P ( 1 + P )
\label{eq:P1}
\end{equation}
and introducing the auxiliary state $\mid \tilde U \rangle$:
\begin{equation}
\mid \tilde U \rangle \equiv t G_0 \mid U \rangle + \frac12 ( 1 + P ) V_4^{(1)} G_0 ( t G_0 +1) \mid U \rangle
\end{equation}
one gets from Eq.~(\ref{Eq:start})
\begin{equation}
\mid U \rangle = ( 1 + P ) j_\mu \mid \Psi_{\rm b} \rangle + P \mid \tilde U \rangle \; .
\end{equation}
Then the Faddeev-like equation for the state $\mid \tilde U \rangle$  is
\begin{eqnarray}
\mid \tilde U \rangle  &=&
\left( t G_0 + \frac12 ( 1 + P ) V_4^{(1)} G_0 ( t G_0 +1) \right) \nonumber \\
&*&( 1 + P ) j_\mu \mid \Psi_{\rm b} \rangle \nonumber \\
&+& ( t G_0 P + \frac12 ( 1 + P ) V_4^{(1)} G_0 ( t G_0 +1) P
\mid \tilde U \rangle ,
\label{eq:Utilde}
\end{eqnarray}
and the transition amplitudes are
\begin{eqnarray}
N_\mu^{\rm Nd} =
\langle \phi_{1} \mid  ( 1 + P ) \mid
j_\mu \mid \Psi_{\rm b} \rangle +
\langle \phi_{1} \mid  P \mid \tilde U \rangle
\label{eq:Nnew}
\end{eqnarray}
and
\begin{eqnarray}
N_\mu^{\rm 3N} &=&
\langle \phi_{3N} \mid ( 1 + P )
j_\mu \mid \Psi_{\rm b} \rangle \nonumber \\
&+&
\langle \phi_{3N} \mid t G_0 ( 1 + P )
j_\mu \mid \Psi_{\rm b} \rangle \nonumber \\
&+&
\langle \phi_{3N} \mid P \mid \tilde U \rangle +
\langle \phi_{3N} \mid t G_0 P \mid \tilde U \rangle .
\label{eq:N3new}
\end{eqnarray}
 
 
In the case when only the NN interaction is used ($V_4^{(1)}=0$)
one gets from Eq.~(\ref{Eq:start})
\begin{equation}
\mid U \rangle = \mid \chi \rangle + Pt_1G_0 \mid U \rangle \;. \label{eq17}
\end{equation}
Using two consequences of the identity~(\ref{eq:P1}):
\newline
$\mid \chi \rangle = \frac{1}{2} P \mid \chi \rangle$ and $2=P(P-1)$
Eq.~(\ref{eq17}) can be rewritten as
\begin{equation}
\mid U \rangle = \frac12 P [ \mid \chi \rangle + t_1G_0P (P-1) \mid U \rangle ]\;. \label{eq18}
\end{equation}
This is equivalent to
\begin{equation}
\mid U \rangle =\frac12 P \mid U\ptrzy \rangle \;, \label{eq19}
\end{equation}
where $\mid U\ptrzy \rangle$ fulfills
\begin{equation}\label{met9}
\mid U\ptrzy \rangle = \mid \chi \rangle + t_1G_0P \mid U\ptrzy \rangle \;.
\end{equation}
This equivalence can be easily seen by iterating Eq.~(\ref{eq18}) and Eq.~(\ref{met9}).
Therefore one obtains from Eq.~(\ref{neweq8}) and Eq.~(\ref{eq19})
\begin{equation}
N^{Nd}_{\mu} = \frac{1}{2} \langle \phi_1 \mid P \mid U\ptrzy \rangle \;.\label{r11}
\end{equation}
The transition amplitude for the three-body photodisintegration is given by~\cite{skib2b}
\begin{equation}
N^{3N}_{\mu} = \frac{1}{2} \langle \phi_{3N} \mid P \mid U\ptrzy \rangle +
\frac{1}{2} \langle \phi_{3N} \mid t G_0 P \mid U\ptrzy \rangle \;. \label{r11s}
\end{equation}                
 
The results based on Eqs.~(\ref{r11}) and~(\ref{r11s})will be denoted as "method $2_{NN}$."
In both methods  $2_{NN+3NF}$ and  $2_{NN}$ one meets the action of the permutation
operator P onto
matrix elements of the P-operator from the previous iteration.
In the actual numerical implementation the identity~(\ref{eq:P1}) is fulfilled only approximately.
Since we work in a partial wave decomposition,
we can take into account only a finite number of partial waves.
In order to fulfill the identity~(\ref{eq:P1}) a high number of partial waves has to be used.
However, also the number
of necessary two-body channels increases with the value of total momentum of the two-body subsystem.
In consequence, this requires a large amount of memory and computing time.
We have found that the worst convergence occurs in the case of method $2_{NN}$ and one
needs to use two-body channels with $j_{max} =5$ during the iteration of Eq.(~\ref{met9}).
It is due to the lack of the t-matrix in the driving term of this equation.
In all other methods
the t-matrix, which is most active in the lower channels,
reduces the influence of higher two-body channels. Therefore even
using the identity~(\ref{eq:P1}) one can restrict the number of partial waves.
 
For all the methods described above, one obtains the transition amplitude $N^{rad}_{\mu}$
for the radiative capture
using time reversal in the two-body photodisintegration amplitude $N^{Nd}_{\mu}$.
 
\subsection{Methods $3_{NN+3NF}$ and $3_{NN}$}
Another possibility to obtain the transition amplitude $N^{rad}_{\mu}$ is to calculate it directly for
the radiative capture process. Then methods used for the elastic Nd scattering can be used
where the initial Nd channel state occurs in  the driving term of the corresponding equation
~\cite{Golak.pdc.2000}. This makes a clear difference
to the previous methods. Then the matrix elements of the nuclear current comes in the last
stage of the calculations
after the solution of the Faddeev  equation is obtained.   
 
In this case we calculate directly the transition amplitude for the radiative Nd-capture
\begin{equation}
N^{rad}_{\mu}= \langle \Psi_b \mid j_{\mu} \mid \Psi_{Nd}^{(+)} \rangle \;.
\end{equation}
The Faddeev component $\mid \psi_1 \rangle $ forming $\mid \Psi_{Nd}^{(+)} \rangle$ as
\begin{equation}
\mid \Psi_{Nd}^{(+)} \rangle = (1+P) \mid \psi_1 \rangle
\end{equation}
is given via
\begin{equation}
\mid \psi_1 \rangle = \mid \phi_1 \rangle +G_0 \tilde T \mid \phi_1 \rangle \;,
\end{equation}
where $\tilde{T} \mid \phi_1 \rangle$ obeys
\begin{eqnarray}
\tilde{T} \mid \phi_1 \rangle &=& t P \mid \phi_1 \rangle +
(1 + t G_0) V_4^{(1)} (1 + P) \mid \phi_1 \rangle \nonumber \\
&+& t P G_0 \tilde{T} \mid \phi_1 \rangle \nonumber \\
&+&
(1 + t G_0) V_4^{(1)} (1 + P) G_0 \tilde{T} \mid \phi_1 \rangle \label{eqtphi}
\end{eqnarray}
 
Solving Eq.~(\ref{eqtphi}) one gets the amplitude $\tilde{T} \mid \phi_1 \rangle$.
The next step in the numerical implementation is to apply the free propagator $G_0$ and to
obtain $\mid \psi_1 \rangle$.
Finally $\langle \Psi_b \mid j_{\mu}$ is acted on obtaining the transition
amplitude $N^{rad}_{\mu}$.
This method we denote by "method $3_{NN+3NF}$."
 
In the absence of the 3N force Eq.~(\ref{eqtphi}) simplifies to
\begin{equation}
\tilde{T} \mid \phi_1 \rangle = t P \mid \phi_1 \rangle + t P G_0 \tilde{T} \mid \phi_1 \rangle \label{eq22}
\end{equation}
and the corresponding result will be denoted as "$3_{NN}$."
In our numerical implementation we always calculate
the matrix elements of the current operator in the frame in which the photon
momentum is parallel to $z$-axis.
To obtain the transition amplitude for each final angle between the outgoing photon
and the beam direction we have to adjust the corresponding initial deuteron or proton
beam direction. This demands the repetition of the iteration of Eq.~(\ref{eqtphi}) in each
case.
One can avoid this by calculating the matrix elements of the current operator in a
general frame.                 

Of course, the methods $3_{NN+3NF}$ and $3_{NN}$ by construction can be used only for
the Nd-radiative capture and the two-body photodisintegration.
 
\section{Results}
\label{secIII}
 
We would like to present the quality of our methods for pd-capture
for six exemplifying observables: the nucleon and deuteron vector analyzing powers $A_Y(N)$ and
$A_Y(d)$, the differential cross sections and the tensor analyzing powers
$A_{YY}$, $A_{XX}$  and $T_{21} \equiv \frac{-1}{\sqrt{3}} A_{XZ}$.
 
\begin{figure}[h!]
\leftline{\mbox{\epsfysize=180mm \epsffile{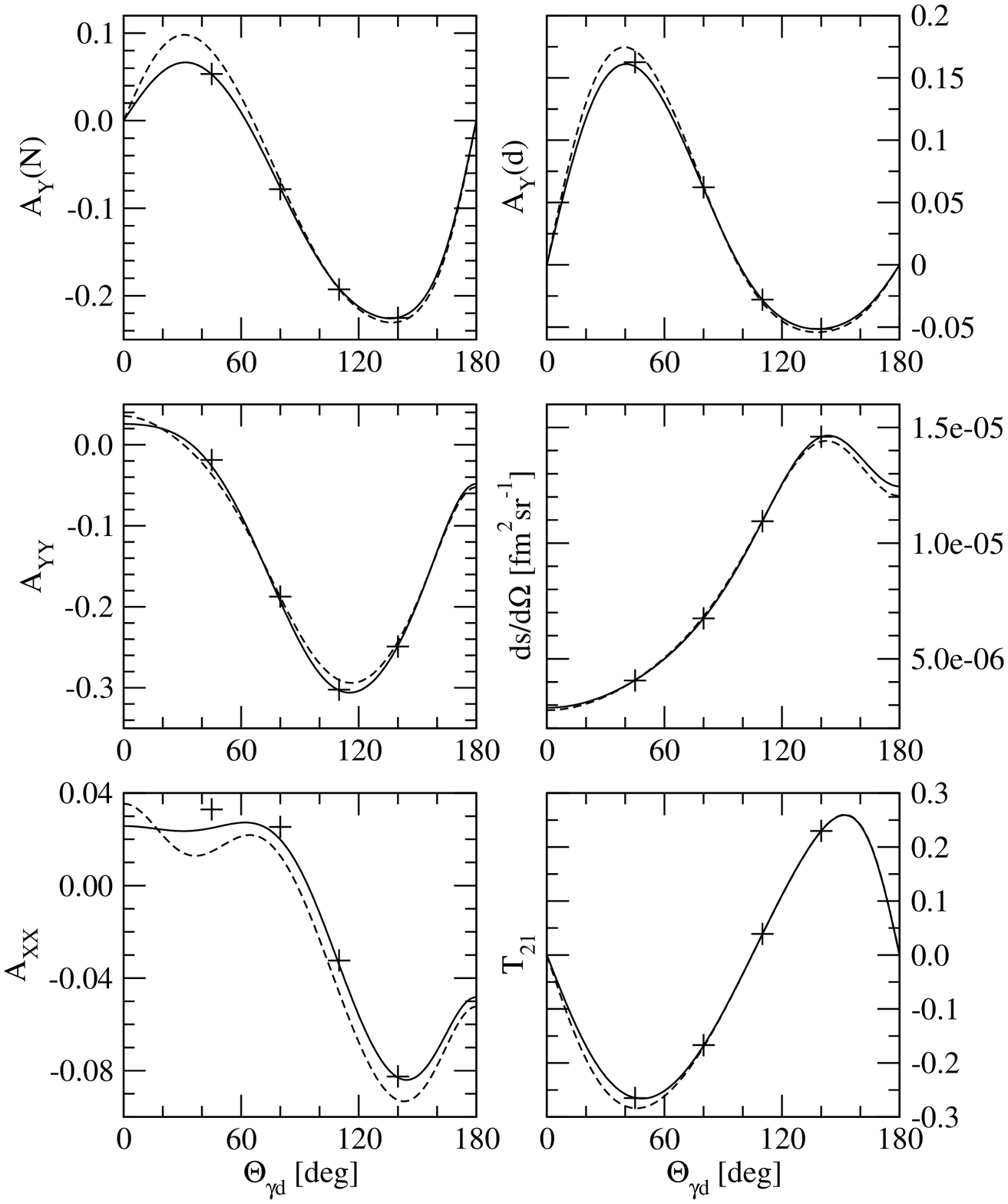}}}
\caption[ ]
{
The comparison of the three methods using the AV18 interaction at
the deuteron laboratory energy E$_d$=300 MeV.
The solid line represent predictions based on method $1_{NN}$,
the dashed one on method $2_{NN}$
(here j$_{\rm max}$=5) and the crosses are for method $3_{NN}$ (see text).
}
\label{fig1}
\end{figure}

Predictions based on the AV18 interaction for these observables
for pd capture are presented in Fig.~\ref{fig1}.
The incoming deuteron laboratory energy is $E_d=300$ MeV. This  is equivalent
to a photon laboratory energy $E_\gamma \approx$106  MeV in the photodisintegration process.
The iteration of the Faddeev-like equations uses a lot of computer power. Thus in the case of
method $3_{NN}$  which in our numerical implementation, as mentioned above, demands a separate
solution of Eq.(~\ref{eq22}) for each angle of the outgoing photon
we present only a few points (denoted by crosses).
We see that the methods $1_{NN}$ and $3_{NN}$ agree
nicely in all cases, except for the tensor analyzing power $A_{XX}$ at small angles.
This agreement is obtained taking into account in both methods the partial waves with total
angular momenta in the two-body subsystem up to $j=3$. The method $2_{NN}$ demands much more
partial waves (up to $j=5$) and still there are more angular ranges where
the predictions of method $2_{NN}$ differ from those of methods $1_{NN}$ and $3_{NN}$.
As mentioned above this is due to the lack of the t-matrix in the driving term.    

\begin{figure}[h!]
\leftline{\mbox{\epsfysize=180mm \epsffile{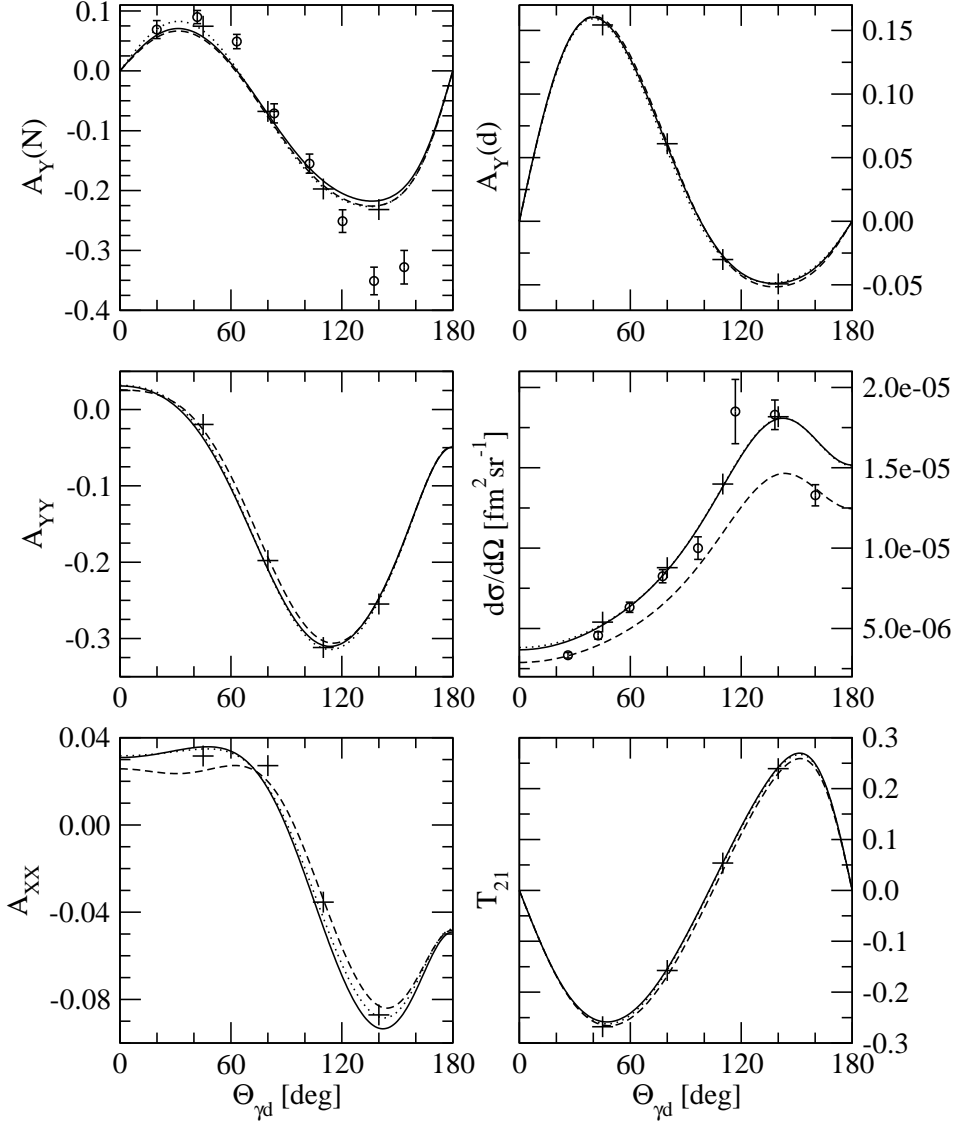}}}
\caption[ ]
{
The comparison of the three methods using the AV18+UrbanaIX interaction at
the deuteron laboratory energy E$_d$=300 MeV, j$_{\rm max}$=3, J$_{\rm max}=\frac{15}{2}$.
The solid line represent predictions based on method $1_{NN+3NF}$,
the dotted one on method $2_{NN+3NF}$
and the crosses are for method $3_{NN+3NF}$.
The dashed line represents predictions of method $1_{NN}$,
based only on the AV18 interaction. Data ($\circ$) are from~\cite{pickar}. 
}
\label{fig2}
\end{figure}

\begin{figure}[h!]
\leftline{\mbox{\epsfysize=180mm \epsffile{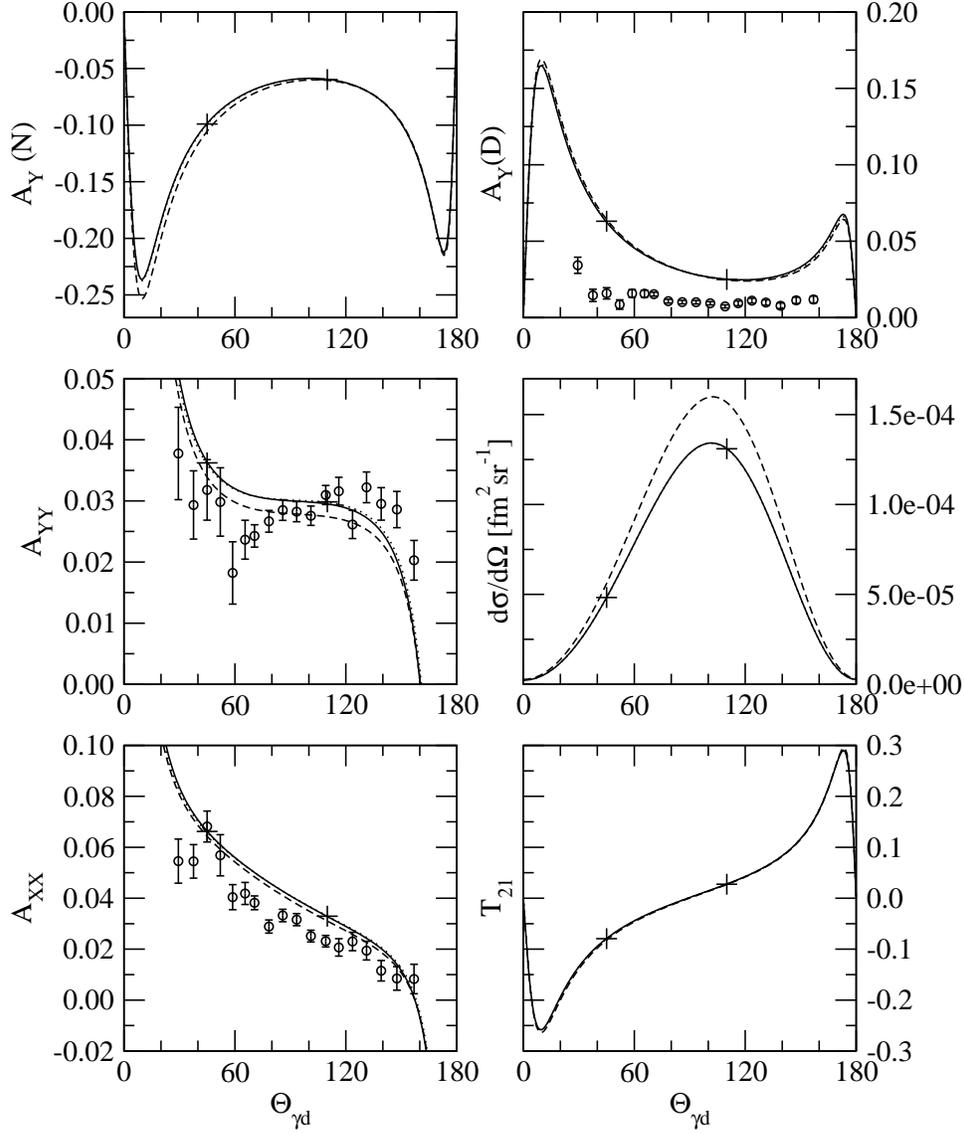}}}
\caption[ ]
{
The comparison of the three methods using the AV18+UrbanaIX interaction at the deuteron laboratory energy
E$_d$=17.5 MeV, j$_{\rm max}$=3, J$_{\rm max}=\frac{15}{2}$.
Curves and crosses as in Fig.~\ref{fig2}. Data ($\circ$) are from~\cite{sagara}.    
}
\label{fig3}
\end{figure}

A nice agreement is obtained when comparing results of methods $1_{NN+3NF}$, $2_{NN+3NF}$ and $3_{NN+3NF}$.
This is shown in Fig.~\ref{fig2} for
the same deuteron energy, E$_d$=300 MeV, and for a much lower one, E$_d$=17.5 MeV,
in Fig.~\ref{fig3}.
The cross section is especially insensitive to the used me\-thod.
In addition to predictions based on the AV18 and the Urbana IX forces in Figs.~\ref{fig2} and ~\ref{fig3}
we present also results based on the AV18 interaction alone.
As we see, except for the cross section, the three nucleon force effects are very small.
Comparison to the Sagara data~\cite{sagara} at E$_d$=17.5 MeV shows a reasonable
agreement for the tensor analyzing powers
$A_{XX}$ and $A_{YY}$ and the known disagreement for vector analyzing
power $A_Y(d)$~\cite{Golak.pdc.2000}.
Comparing our predictions to the Pickar data~\cite{pickar} at E$_d$=300 MeV we see
that for the proton analyzing power $A_Y(N)$ we agree with the data at the lower and
middle angles, while at the higher
angles all theoretical predictions are above the data. In the case of the cross section
differences are much smaller, however the theoretical predictions seem to be flatter than data.
It is also very apparent that the inclusion of the Urbana IX 3NF improves the
description of the data.
Only two experimental points at very small and at very large angles are in
better agreement with the pure NN force prediction.
 
As is well known, for low energies the 3NF contributes mainly in the 3N bound state,
while for higher energies the 3NF effects are seen also in the continuum.
This is the reason why we present results also for a relatively high energy (E$_d$=300 MeV).
We also note that the agreement between all three methods is slightly better
for lower energy, especially for the
deuteron tensor analyzing powers (e.g. A$_{XX}$).
 
\begin{figure}[h!]
\leftline{\mbox{\epsfysize=180mm \epsffile{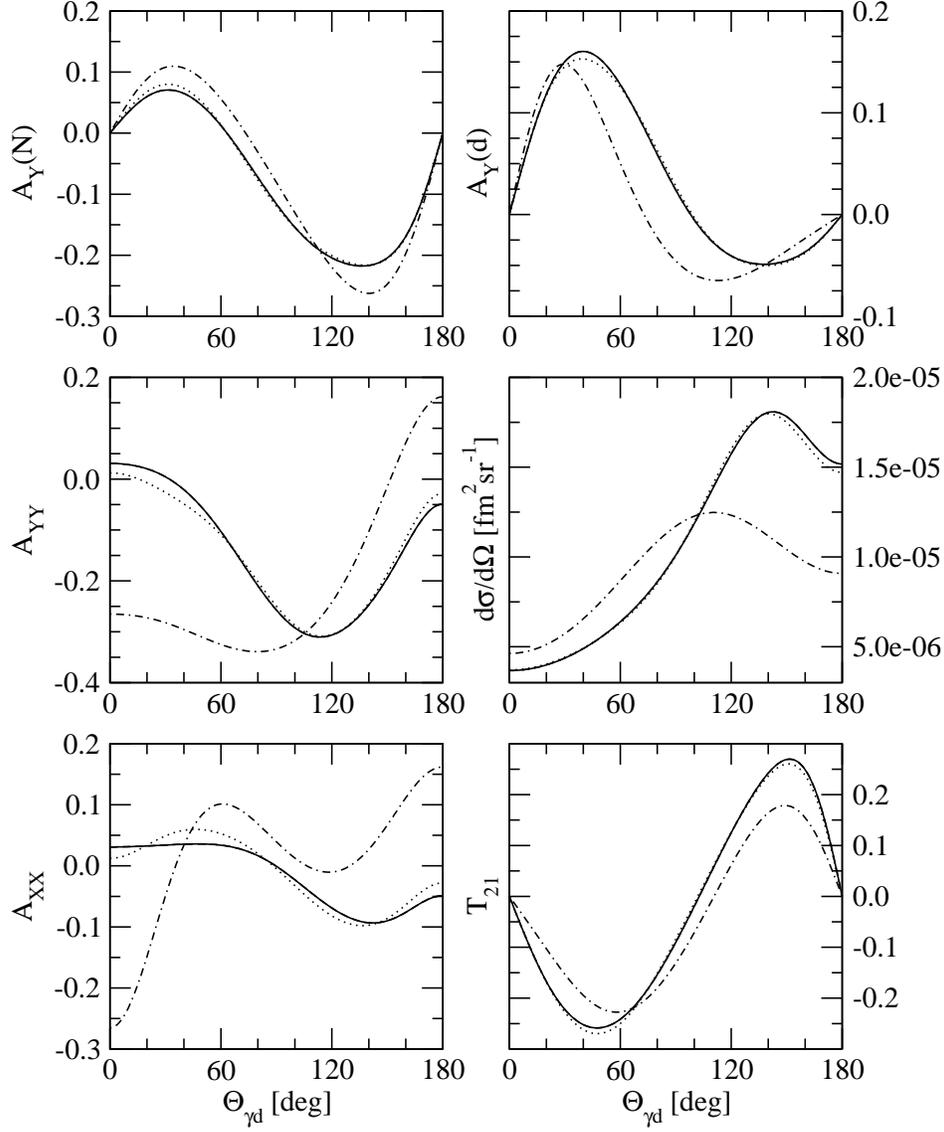}}}
\caption[ ]
{
The convergence in the number of three-body partial waves for the method $1_{NN+3NF}$
using the AV18+UrbanaIX interaction.
E$_d$=300 MeV, j$_{\rm max}$=3 fixed, J$_{\rm max}=\frac{3}{2}$ (dash-dotted),$\frac{7}{2}$ (dotted),$\frac{15}{2}$ (dashed),
$\frac{19}{2}$ (solid line). Dashed and solid lines are indistinguishable.      
}
\label{fig4}
\end{figure}

\begin{figure}[h!]
\leftline{\mbox{\epsfysize=180mm \epsffile{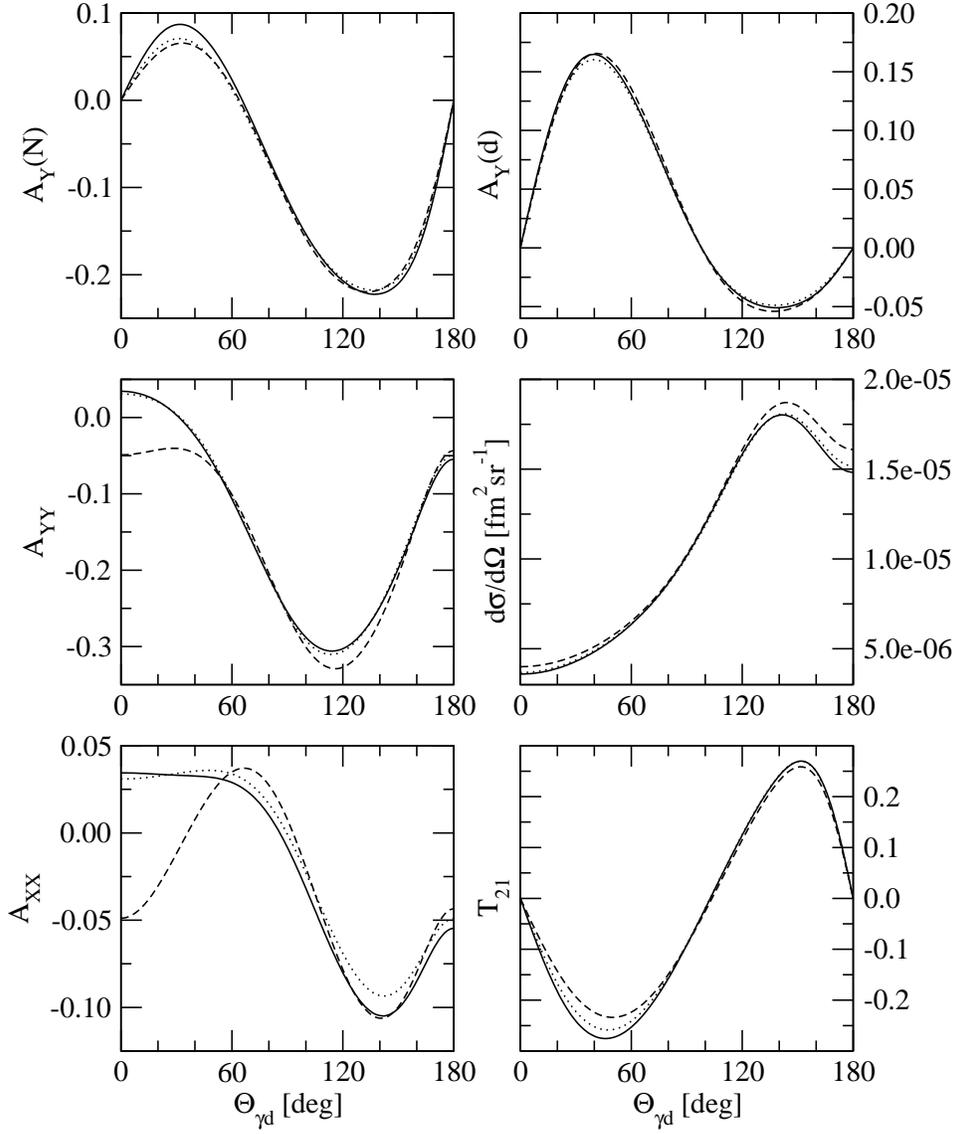}}}
\caption[ ]
{
The convergence in the number of two-body partial waves for method $1_{NN+3NF}$
using the AV18+UrbanaIX interaction.
E$_d$=300 MeV, J$_{\rm max}$=$\frac{15}{2}$ fixed, j$_{\rm max}=2$ (dashed),
3 (dotted) and 4 (solid line).
}        
\label{fig5}
\end{figure}

The next two figures show dependences of our results on the number of partial waves for the
example of method $1_{NN+3NF}$.
In Fig.~\ref{fig4} we show the convergence of the
predictions in the total angular momentum $J_{max}$ of the three-body system.
While using only channels with $J_{max} = \frac{3}{2}$ is absolutely insufficient,
using channels with $J_{max} = \frac{7}{2}$ is very close to the final prediction with
$J_{max} = \frac{15}{2}$. Allowing the transitions to higher total angular momenta states
(up to $J_{max} = \frac{19}{2}$) does not change the predictions in a perceptible manner.
A similar picture appears for the convergence in the number of partial waves used in the
two-body subsystem
(see Fig.~\ref{fig5}). While using only channels with total angular momenta in the two-body
subsystem up to $j=2$ is far from the predictions with $j=4$, there is only a small difference between
the predictions with maximal values of the two-body total angular momentum $j=3$ and $j=4$.
However, the method $2_{NN}$ (not shown in Fig.~\ref{fig5}) requires at least $j_{max}=5$.   

\begin{figure}[h!]
\leftline{\mbox{\epsfysize=60mm \epsffile{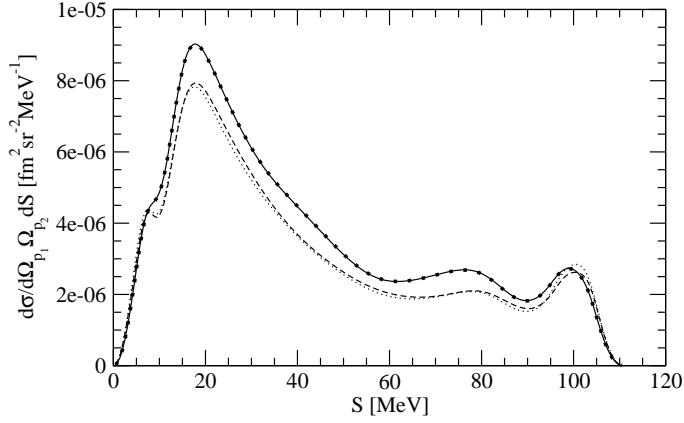}}}
\caption[ ]
{
The differential cross section for three-body photodisintegration
at E$_\gamma$=100 MeV at the two proton angles:
$\Theta_1=10^\circ$, $\Phi_1=0^\circ$, $\Theta_2=10^\circ$, $\Phi_2=0^\circ$.
The dotted, dashed, solid and thick dotted curves represent methods
$1_{NN}$, $2_{NN}$, $1_{NN+3NF}$, $2_{NN+3NF}$,
respectively.
}
\label{fig6}
\end{figure}

\begin{figure}[h!]
\leftline{\mbox{\epsfysize=60mm \epsffile{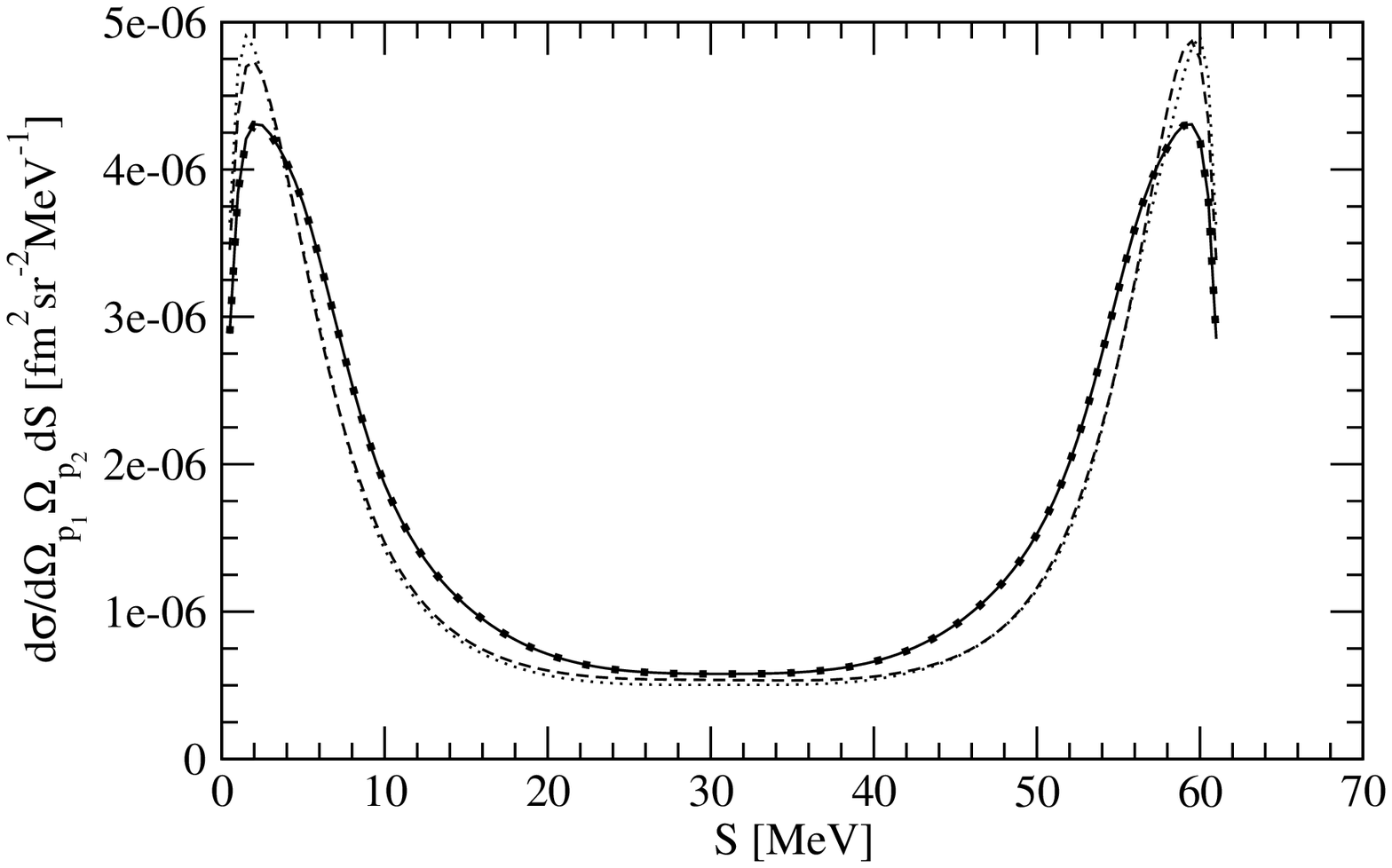}}}
\caption[ ]
{
The differential cross section for three-body photodisintegration at E$_\gamma$=100 MeV at protons angles:
$\Theta_1=90^\circ$, $\Phi_1=0^\circ$, $\Theta_2=90^\circ$, $\Phi_2=90^\circ$.
Curves as in Fig.~\ref{fig6}.  
}
\label{fig7}
\end{figure}

\begin{figure}[h!]
\leftline{\mbox{\epsfysize=60mm \epsffile{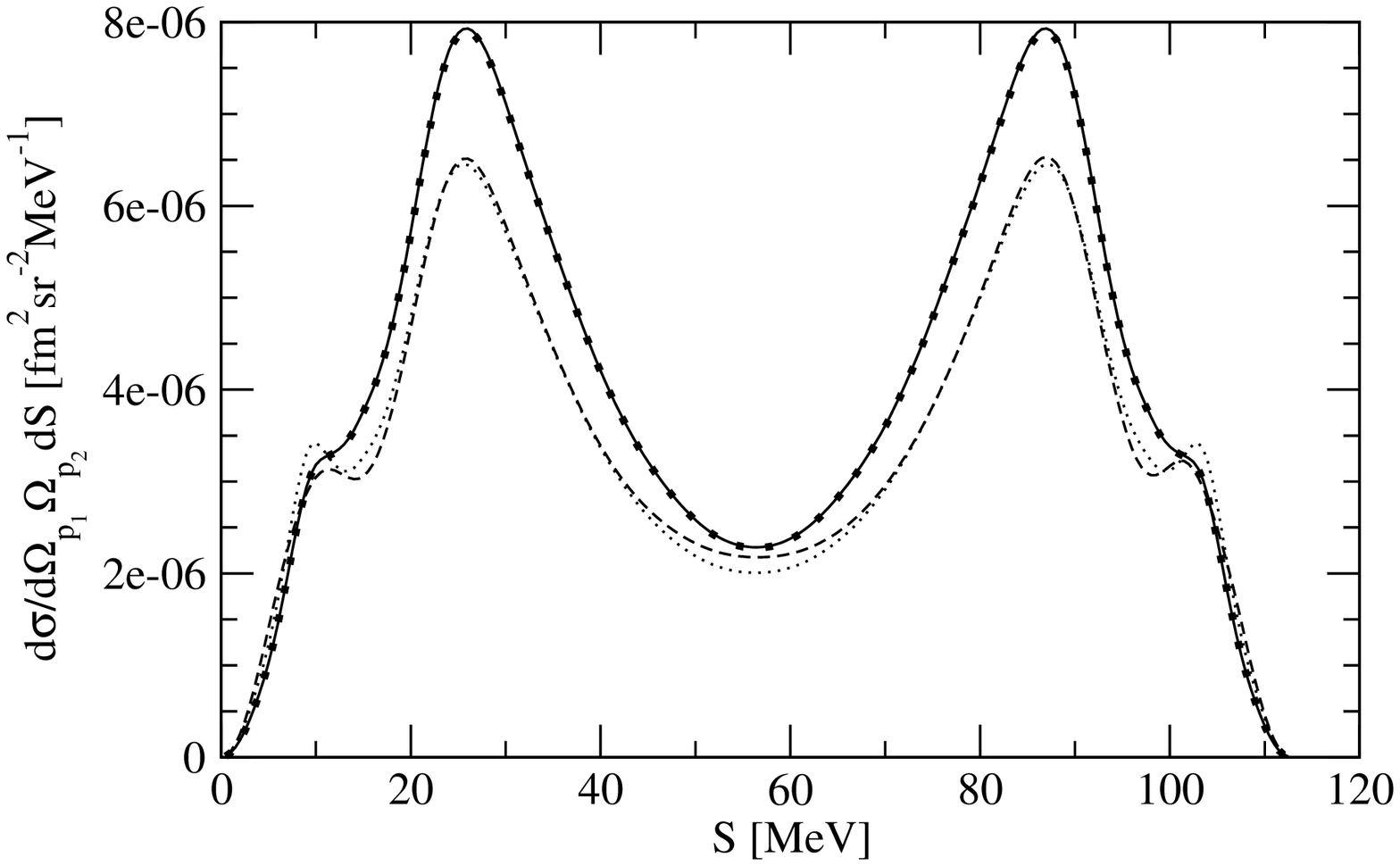}}}
\caption[ ]
{
The differential cross section for three-body photodisintegration at E$_\gamma$=100 MeV at protons angles:
$\Theta_1=90^\circ$, $\Phi_1=0^\circ$, $\Theta_2=90^\circ$, $\Phi_2=180^\circ$.
Curves as in Fig.~\ref{fig6}.  
}
\label{fig8}
\end{figure}

\begin{figure}[h!]
\leftline{\mbox{\epsfysize=60mm \epsffile{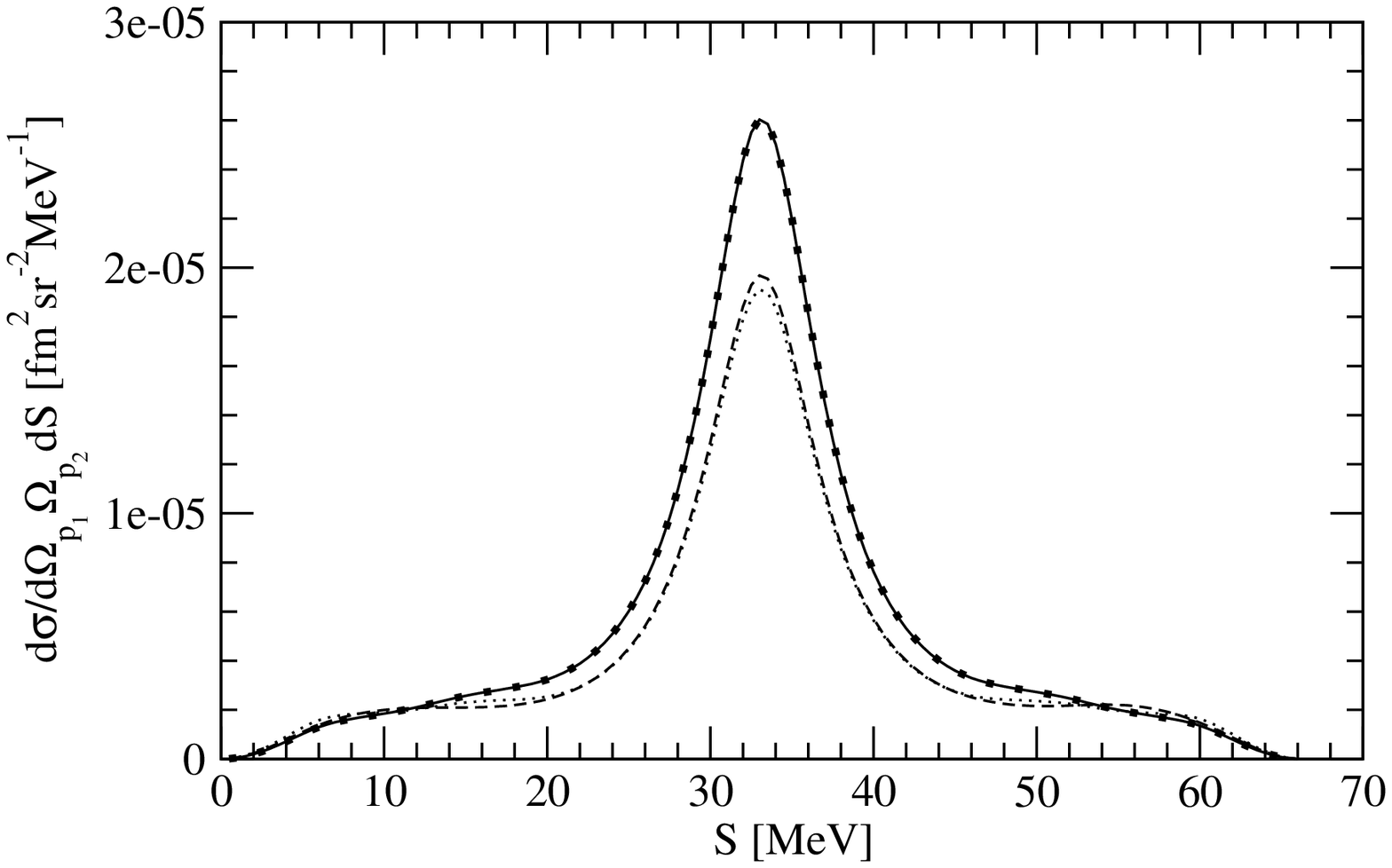}}}
\caption[ ]
{
The differential cross section for three-body photodisintegration at E$_\gamma$=100 MeV at protons angles:
$\Theta_1=90^\circ$, $\Phi_1=55^\circ$, $\Theta_2=90^\circ$, $\Phi_2=65^\circ$.
Curves as in Fig.~\ref{fig6}.   
}
\label{fig9}
\end{figure}

Finally, we would like to compare predictions for the three-body
photodisintegration of $^3$He. In Figs.~\ref{fig6}-\ref{fig9} we present examples of exclusive differential
cross sections for different kinematical configurations at E$_\gamma$=100 MeV, given as a function of the S-curve arc-length.
The two protons are measured
under different polar and azimuthal angles $\Theta$, $\Phi$. In all cases we see that there is an excellent agreement
between the predictions based on methods $1_{NN+3NF}$ and $2_{NN+3NF}$. They are
represented by solid and thick dotted curves, respectively.
In both cases we show results
with channels up to $j_{\max} \leq 3$ and $J_{\max} \leq \frac{15}{2}$.
As we checked for a large number of configurations (about 300 000) the
differences remain for all cases below 1\%.
In the case of the predictions based
on NN interaction only, the differences are a bit bigger - for the majority of the
configurations they are below 5\%.
In that case channels with $j_{\max} \leq 4$ were used for the method $1_{NN}$ (dotted line)
and with
$j_{\max} \leq 5$  for the method $2_{NN}$ (dashed line).
Again, the reason are the different forms of the driving term in the two methods.   

\section{Summary}

Different formulations of Faddeev-like equations for pd-capture (equivalent
to two-body photodisintegration of the 3N bound state) and for three-body photodisintegration
of the 3N bound state
have been investigated.
This is important to guarantee reliable and well converged theoretical results for modern
NN and 3N forces. Such tools allow an unambiguous test of the dynamics when compared to data.
This study is especially timely since a new approach to nuclear forces and currents based on
effective field theory constrained by chiral symmetry is under vivid development~\cite{chiral1}
-\cite{chiral2}.
As we demonstrated all methods show a good agreement. Also the computer time needed is very
similar. Therefore, none of them is preferable, but all are equally useful.
The comparison, however, was a very important internal benchmark of the methods.
Only the method $2_{NN}$ is much slower, since there more two-body partial waves have to be used.
For that reason we have not checked another fourth possible formulation with the same driving term
as in the method $2_{NN}$, but
also three nucleon interaction in the integral kernel~\cite{skib2b}.
The reason for the slow convergence is
that the permutation operator from the integral kernel acts onto
the permutation operator from the driving term and
one has to use a much bigger number of
partial waves to obtain converged results.
 
We can conclude that now different methods are available, which can be considered to be
very reliable and the achieved results document the numerical accuracy up to the order of
a few percent slightly dependent on the observable. At the low energies the accuracy is
much better. In addition this can be improved if smaller experimental errors in the future
will require that.                                

\acknowledgements
This work was supported by
the Polish Committee for Scientific Research
under Grants No. 2P03B0825
and by US DOE under grants Nos. DE-FC02-01ER41187 and
DE-FG02-00ER41132.
The numerical calculations have been performed
on the Cray T90, SV1 and IBM Regatta p690+ of the NIC in J\"ulich, Germany.


\vspace{2.cm} 

\vspace{2.cm}

\end{document}